\documentstyle[prb,epsf,aps,multicol]{revtex}

\def\beq{\begin{equation}}
\def\eeq{\end{equation}}

\def\prb{Phys. Rev. B }
\def\pra{Phys. Rev. A }
\def\pla{Phys. Lett. A }
\def\prl{Phys. Rev. Lett. }
\def\ajp{Am. J. Phys. }
\def\mpl{Mod. Phys. Lett. B }
\def\epl{Euro. Phys. Lett. }
\def\ijmp{Int. J. Mod. Phys. B }
\def\ijp{Ind. J. Phys. }
\def\ibmjrd{IBM J. Res. Dev. }
\def\pjp{Pramana J. Phys. }

\begin{document}

\draft

\title{Aharonov-Bohm oscillations and spin transport in a mesoscopic
  ring with a magnetic impurity}

\author{Sandeep K. Joshi\cite{joshie}, Debendranath
    Sahoo\cite{sahooe} and A. M. Jayannavar\cite{amje} } 
 
\address{Institute of Physics, Sachivalaya Marg, Bhubaneswar 751 005,
  Orissa, India} 

\date{\today}

\maketitle

\begin{abstract} 
  
  We present a detailed analysis of the Aharonov-Bohm (AB)
  interference oscillations manifested through transmission of an
  electron in a mesoscopic ring with a magnetic impurity atom inserted
  in one of its arms. The spin polarization transport is also studied.
  The electron interacts with the impurity through the exchange
  interaction leading to exchange spin-flip scattering. Transmission
  in the spin-flipped and spin-unflipped channels are explicitly
  calculated. We show that the entanglement between electron and
  spin-flipper states lead to a reduction of AB oscillations in spite
  of absence of any inelastic scattering. The spin-conductance
  (related to spin-polarized transmission coefficient) is asymmetric
  in the flux reversal as opposed to the two probe conductance which
  is symmetric under flux reversal. We point out certain limitations
  of this model in regard to the general notion of dephasing in
  quantum mechanics.

\end{abstract}

%% \pacs{PACS Nos.: 03.65.Bz, 05.60.Gg, 34.80.Nz, 42.50.Dv, 
%%                  42.50.Md, 75.30.Et, 75.30.Hx}

\begin{multicols}{2}

%%\section{Introduction}
  
\begin{figure}
\protect\centerline{\epsfxsize=2.5in \epsfbox{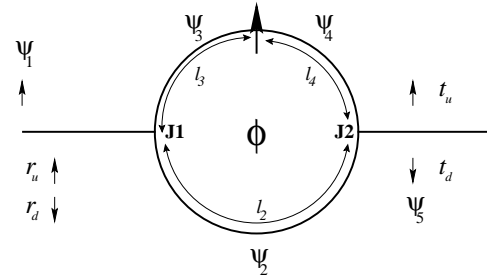}}
\caption{Mesoscopic ring with Aharonov-Bohm flux $\phi$ threading
  through the center of the ring and a magnetic impurity in one arm of
  the ring.}
\label{ring}
\end{figure}

  Quantum transport in open mesoscopic systems has attracted
  considerable attention in the last two decades
  \cite{imry_book,datta,psd}. In this area, the study of phase
  coherent transmission of electrons in the Aharonov-Bohm (AB) ring
  occupies a prominent place
  \cite{imry_book,datta,psd,webb_ap,gia,wgtr}. Study of dephasing and
  decoherence \cite{sai,ims_pb,butt_prb,pareek,nku} of electrons in
  this geometry is very timely \cite{shtrikman} to understand basic
  issues related to quantum phenomena. By introducing a magnetic
  impurity atom (to be referred to as the spin-flipper, or the
  flipper, for short) in one arm of the ring, one can couple the spin
  of the electron ($\vec{\sigma}$) to the spin of the flipper
  ($\vec{S}$) via the exchange interaction \cite{imry_book,sai}.  This
  leads to scattering of the electron in which the spin state of the
  electron and the impurity is changed without any exchange of energy.
  Additionally, this scattering leads to the entanglement-induced
  reduction of interference pattern \cite{schul}. Let the electron be
  incident from the left reservoir with its spin pointing ``up'' (see
  Fig.~\ref{ring}). The spin of the electron passing through the upper
  arm may or may not be flipped by the flipper. In the case that the
  spin is unflipped, one would expect the usual AB-oscillations of the
  transmission due to interference of the partial waves passing
  through the upper and the lower branches of the ring. However, in
  the case that the spin is flipped, one would think, guided by naive
  intuition, that a path detection has taken place and hence one would
  be led to conclude that the interference pattern for the spin-down
  component would be wiped out. This is true provided we consider only
  two forward propagating partial waves. However, there are infinitely
  many partial waves in this geometry which are to be superposed to
  get the total transmission. These arise due to the multiple
  reflections from the junctions and the impurity site. Consider, for
  example, an incident spin-up particle moving in the upper arm which
  is flipped at the impurity site and gets reflected to finally
  traverse the lower arm before being transmitted. Naturally, this
  partial wave will interfere with the spin-flipped component
  transmitted along the upper arm. This results in non-zero
  transmission for the spin-flipped electron.  Thus on taking into
  account the multiple reflections (more than just two partial waves)
  the presence of magnetic impurity does not lead to "which-path"
  information. However, the presence of magnetic impurity leads to the
  reduction of AB-oscillations\cite{imry_book,sai,schul}.

In this work we study the reduction of amplitude of AB-oscillations
arising due to the flipper. Within the same model we also study
spin-polarized transport. There has been a great deal of interest in
the concept of spintronics \cite{prinz}, wherein, one
manipulates with the spin of the charge carriers. Devices in which the
electron spin stores and transmits information are being studied.
Finally we discuss the limitations of this model (based on the
interaction induced entanglement of quantum states) to the general
understanding of dephasing in quantum systems. We study the problem
using the quantum waveguide theory approach \cite{wgtr,xia} and the
spin degree of freedom of the electron is dealt with in line with Ref.
\onlinecite{ajp}. We consider an impurity consisting of a flipper
capable of existing in M different discrete internal spin states and
located at a particular position on the upper arm of the ring (see
Fig. \ref{ring}). The spin $\vec{\sigma}$ of the electron couples to
the flipper spin $\vec{S}$ via an exchange interaction $-J
\vec{\sigma} \cdot \vec{S} \delta(x-l_3)$. The magnetic flux threading
the ring is denoted by $\phi$ and is related to the vector potential
$A=\phi/l$, $l$ being the ring circumference \cite{xia}.  During
passage of the electron through the ring, the total spin angular
momentum and its $z$-component remain conserved.  We analyze the
nature of spin-up/down and total transmission (reflection)
coefficients. For this we consider the incident electron to be
spin-polarized in the up-direction. We also show that
up/down-transmission coefficients are asymmetric in flux reversal,
i.e., total spin polarization (related to spin conductance
\cite{sarma}) is asymmetric in flux reversal. As expected we find that
the total transmission coefficient which is the sum of spin-up and
spin-down transmission coefficients is symmetric in the flux reversal.
  
Let $l_2$ be the length of the lower arm of the ring and the impurity
atom be placed at a distance $l_3$ from the junction J1, $l_4$ being
the remaining segment length of the upper arm. The various segments of
the ring and its leads are labeled as shown in Fig. \ref{ring} and
the wave functions in these segments carry the corresponding
subscripts. The wave functions in the five segments for a
left-incident spin-up electron can be written as follows:

\begin{eqnarray}
\label{eq:1}
\psi _1&=&(e^{ikx}+r_u e^{-ikx})\chi _m\alpha+\nonumber\\
          &&r_d e^{-ikx}\chi _{m+1}\beta,\nonumber\\
\psi _2&=&(A_u e^{ik_1x}+B_u e^{-ik_2x})\chi _m\alpha+\nonumber\\
          &&(A_d e^{ik_1x}+B_d e^{-ik_2x})\chi _{m+1}\beta,\nonumber\\
\psi _3&=&(C_u e^{ik_1x}+D_u e^{-ik_2x})\chi _m\alpha+\nonumber\\
          &&(C_d e^{ik_1x}+D_d e^{-ik_2x})\chi _{m+1}\beta,\nonumber\\
\psi _4&=&(E_u e^{ik_1x}+F_u e^{-ik_2x})\chi _m\alpha+\nonumber\\
          &&(E_d e^{ik_1x}+F_d e^{-ik_2x})\chi _{m+1}\beta,\nonumber\\
\psi _5&=&t_u e^{ikx}\chi _m\alpha+t_d e^{ikx}\chi _{m+1}\beta.
\end{eqnarray}

\noindent where $k_1=k+(e\phi /\hbar cl)$, $k_2=k-(e\phi /\hbar cl)$,
$k$ is the wave-vector of incident electron. The wavefunction in
Eq.~\ref{eq:1} is a correlated function (entangled state) of the
electron and the impurity spin which takes into account that the
exchange interaction conserves the $z$-component of the total spin.
The subscripts $u$ and $d$ represent ``up'' and ``down'' spin states
of the electron with the corresponding spinors $\alpha$ and $\beta$
respectively (i.e., $\sigma _z\alpha =\frac{1}{2}\alpha$, $\sigma
_z\beta =-\frac{1}{2}\beta$) and $\chi _m$ denotes the wave function
of the impurity \cite{ajp} with $S_z=m$ (i.e., $S_z\chi _m=m\chi _m$).
The reflected (transmitted) waves have amplitudes $r_u$ ($t_u$) and
$r_d$ ($t_d$) corresponding to the ``up'' and ``down'' spin components
respectively.  Continuity of the wave functions and the current
conservation\cite{wgtr,xia,ajp} at the junctions J1 and J2 imply the
following boundary conditions.

\begin{eqnarray}
  \label{eq:4}
  \psi _1(x=0)=\psi _2(x=0),\nonumber\\
  \psi _1(x=0)=\psi _3(x=0),\nonumber\\
  \psi _1^\prime (x=0)=\psi _2^\prime (x=0)+\psi _3^\prime (x=0),\nonumber\\
  \psi _3^\prime (x=l_3)-\psi _4^\prime
          (x=l_3)=G(\vec{\sigma}\cdot\vec{S})\psi _3(x=l_3),\nonumber\\
  \psi _3(x=l_3)=\psi_4(x=l_3),\nonumber\\
  \psi _4(x=l_3+l_4)=\psi _5(x=0),\nonumber\\
  \psi _2(x=l_2)=\psi _5(x=0),\nonumber\\
  \psi _2^\prime (x=l_2)+\psi _4^\prime (x=l_3+l_4)=\psi _5^\prime (x=0).
\end{eqnarray}

\noindent Here $G=2mJ/\hbar ^2$ is the coupling constant indicative of the
``strength'' of the spin-exchange interaction.  The primes denote the
spatial derivatives of the wave functions. Equations (\ref{eq:1})
along with the boundary conditions (\ref{eq:4}) were solved to obtain
the amplitudes $t_u$, $t_d$, $r_u$ and $r_d$. Owing to the large
length of the expressions in the following we confine ourselves to the
graphical interpretation of the results. We have taken the flipper to
be a spin-half object ($M=2$) situated in the upper arm.
Now, depending upon the initial state of the flipper we have
possibility of either spin-flip scattering ($\sigma_z=1/2,~S_z=-1/2$)
or no spin-flip scattering ($\sigma_z=1/2,~S_z=1/2$), as demanded by
the conservation of the total spin and its $z$-component. In the case
of no-spin-flip scattering ($\sigma_z=1/2,~S_z=1/2$) the problem at
hand reduces to that of simple potential scattering from the impurity.
We have set $\hbar=2m=1$ and throughout the value of interaction
strength $G$ is given in dimensionless units. The parameters used for
the analysis are mentioned in the figure captions.

\begin{figure}
\protect\centerline{\epsfxsize=2.5in \epsfbox{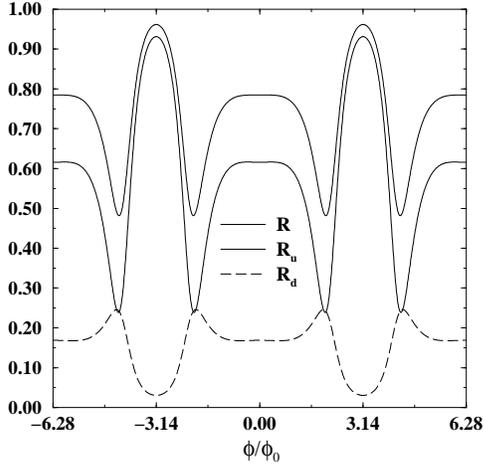}}
\caption{ Plot of total reflection coefficient $R$, spin-up reflection
coefficient $R_u$ and spin-down reflection coefficient $R_d$ for the spin
-flip scattering case. The parameters are $kl=1.0$, $G=10.0$.  }
\label{symm-refl}
\end{figure}

To begin with we take a look at the symmetry properties of the
transport coefficients in spin-flip scattering case where the electron
spin is opposite to the flipper spin. It is worth noting that due to
the presence of spin degree of freedom the problem in hand although
one-dimensional becomes a multichannel problem. Figure \ref{symm-refl}
shows the spin-up reflection coefficient $R_u=|r_u|^2$, spin-down
reflection coefficient $R_d=|r_d|^2$ and total reflection coefficient
$R=R_u+R_d$ as a function of the magnetic flux parameter
$\eta=\phi/\phi_0$, $\phi_0$ being the flux quantum $hc/e$. We clearly
see the AB-oscillations with flux periodicity\cite{webb_ap} of
$2\pi\phi_0$. All three reflection coefficients are symmetric in the
flux reversal as expected on general grounds\cite{butt_ibm}. 

\begin{figure}
\protect\centerline{\epsfxsize=2.5in \epsfbox{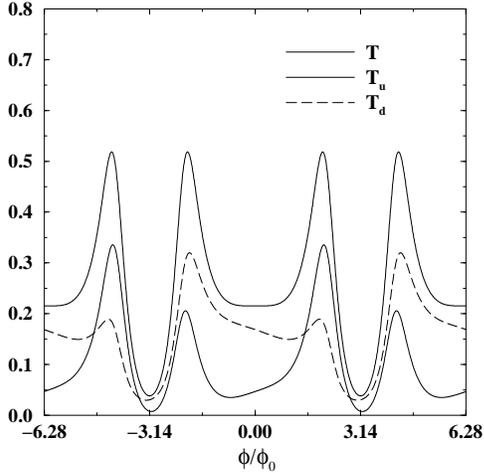}}
\caption{ Plot of total transmission coefficient $T$, spin-up transmission
coefficient $T_u$ and spin-down transmission coefficient $T_d$ for the spin
-flip scattering case. The parameters are $kl=1.0$, $G=10.0$.  }
\label{symm-trans}
\end{figure}

\begin{figure}
\protect\centerline{\epsfxsize=3.2in \epsfbox{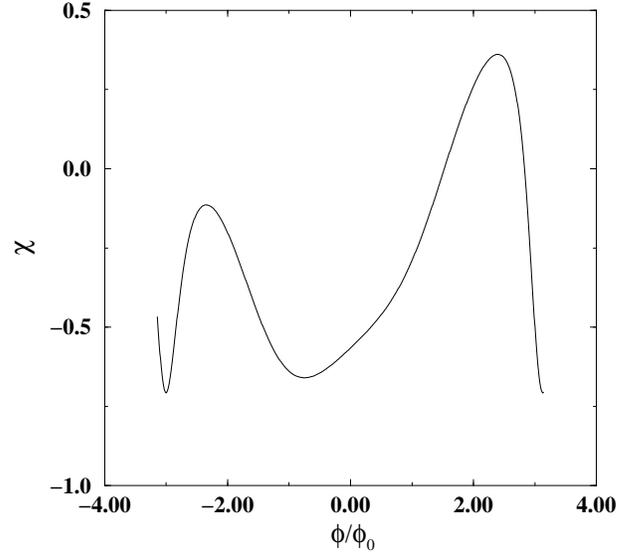}}
\caption{ Spin polarization ($\chi$) as a function of the flux $\phi$
  for interaction strength $G=10.0$. }
\label{sppol}
\end{figure}

In Fig.  \ref{symm-trans} we plot the spin-up transmission coefficient
$T_u=|t_u|^2$ (thin line), spin-down transmission coefficient
$T_d=|t_d|^2$ (dashed line) and total transmission coefficient
$T=T_u+T_d$ (thick line) versus $\eta$. It unambiguously shows that
though the total transmission $T$ (related to the two-terminal
electrical conductance) is symmetric in flux reversal the spin-up
$T_u$ and spin-down $T_d$ components are asymmetric under flux
reversal. These transmission coefficients show AB-oscillations with
flux periodicity of $2\pi\phi_0$.  We have verified this behavior of
the reflection and transmission coefficients for various values of
wave vector $k$ of the incident electron and impurity strength $G$.
These observations are consistent with the reciprocity relations for
transport in multichannel systems \cite{butt_ibm} and are a
consequence of the general symmetry properties of the Hamiltonian
\cite{datta}.  The transmission coefficient at flux $\phi$ for the
case when the incident particle is spin-up and the impurity is
spin-down is equal to the transmission coefficient for the case when
incident particle is spin-down and impurity is spin-up but the flux
direction is reversed.  For the spin-polarized transport the total
polarization $T_u-T_d$ is related to the spin-conductance
\cite{sarma}. The above symmetry properties imply that the
spin-conductance is asymmetric under the flux reversal. This can be
easily noted from Fig.~\ref{sppol}. In the figure we have plotted the
variation of spin polarization $\chi = (T_u-T_d)/T$ as a function of
the magnetic flux $\phi$. This spin-polarization can be experimentally
measured by using the well known spin-valve (magnetic valve or filter)
effect \cite{prinz}. In this method ferromagnetic metal pads are used
at the junctions between the sample and the reservoirs. Using a
ferromagnetic metal contact one can inject spin-polarized electrons.
The transmitted electron intensity or current can be measured by
tuning the magnetization axis of the ferromagnetic contact at the
drain. This contact acts as a spin-polarizer. The current depends on
the relative angle of polarization between the transmitted electrons
and the ferromagnetic metal contact at the drain (well-known
$\cos^2(\theta/2)$ dependence)~\cite{prinz}. Our actual system may
comprise of a single channel clean metallic ring and by doping it with
a paramagnetic impurity atom~\cite{ajp} (or impurities~\cite{dieny})
which has a virtual state in the continuum.  It is also possible to
use a quantum dot with one excess electron in one of the arm to
replace the role of a magnetic impurity.  It should be noted that at
zero temperature the total electrical and spin conductances are to be
calculated by summing up with equal weight-age the total transmission
coefficients for all the four cases, i.e., $\sigma_z=\pm 1/2$ and
$S_z=\pm 1/2$.

\begin{figure}[t]
\protect\centerline{\epsfxsize=3in \epsfbox{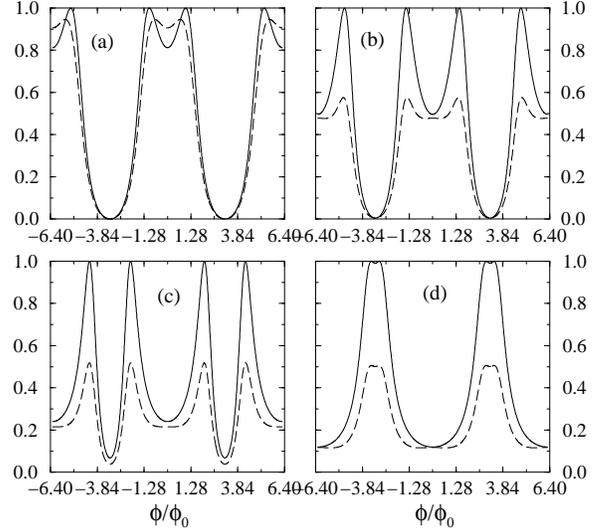}}
\caption{ Amplitude of AB oscillations or interference fringe visibility
  for the two cases of SFS and NSFS for different strengths of the
  exchange interaction. In all four cases $l_2/l=0.5$, $l_3/l = l_4/l
  = 0.25$ and $kl = 1.0$. The values of coupling strength $G$ are (a)
  $G = 1.0$, (b) $G = 5.0$, (c) $G = 10.0$ and (d) $G = 15.0$.}
\label{visib}
\end{figure}

As discussed in the introduction, due to multiple reflections the
presence of a spin-flipper in one arm does not lead to "which-path"
information.  This would have implied the complete blocking of
spin-down transmission.  In contrast we clearly observe the
AB-oscillations for the case of $T_d$ originating from multiple
reflections. We now address the question of partial loss of
interference due to the spin-flipper. In Fig. \ref{visib} we have
plotted the total transmission coefficient $T=T_u+T_d$ for the
spin-flip scattering (SFS) case , and $T=T_u$ ($T_d=0$) for the no
spin-flip scattering (NSFS) case for different parameters as indicated
in the figures \ref{visib}(a-d).  As expected $T$ exhibits AB
oscillations which are periodic in flux with a period $\phi_0$ and
they are symmetric under flux reversal.  It is interesting to note,
however, that the interference fringe visibility (or the magnitude of
amplitude of AB oscillations) for the SFS case is always smaller than
that for the case of NSFS. This clearly indicates partial decoherence.

\begin{figure}[t]
\protect\centerline{\epsfxsize=3in \epsfbox{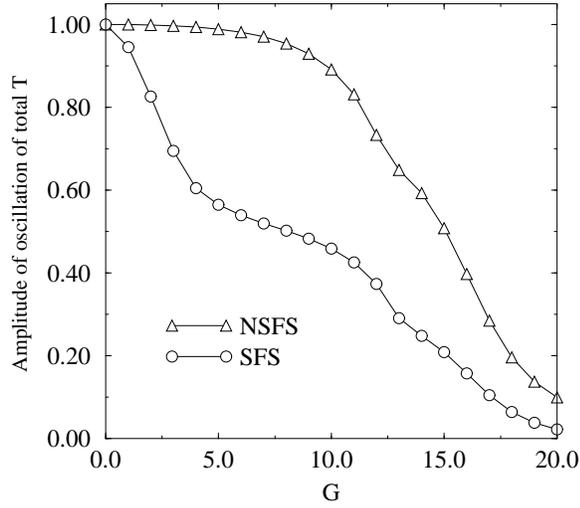}}
\caption{ Variation of Amplitude of AB oscillations with increasing
strength $G$ of spin-flipper for the case of asymmetrically placed
flipper. $l_3/l=0.15$, $l_4/l=0.35$ and $kl=1.0$.}
\label{asymm-amp}
\end{figure}

\begin{figure}
\protect\centerline{\epsfxsize=2.5in \epsfbox{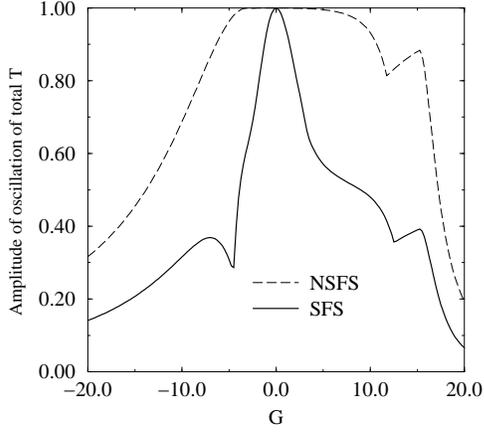}}
\caption{ Variation of amplitude of oscillation of total transmission
coefficient with the interaction strength for the two cases of flip
and no-flip scattering. The parameters are $kl=1.0$. }
\label{ampli-reduc}
\end{figure}

To quantify the decoherence, we calculate the amplitude of AB
oscillations by taking the difference between the maximum and the
minimum of total transmission coefficient as a function of flux $\phi$
over one period of the oscillation. A plot of the variation of the
amplitude of oscillation of total transmission $T$ with the
interaction strength $G$ for the two cases, no spin-flip scattering
(NSFS: $S=1/2~m=1/2$) and spin-flip scattering (SFS: $S=1/2~m=-1/2$),
are shown in figures Fig.~\ref{ampli-reduc} and \ref{asymm-amp}. The
two figures correspond to two different locations of the impurity in
the upper arm of the ring. Other parameters are mentioned in the
respective figure captions. Note, however, the signature of loss of
interference is that the amplitude of AB oscillation of transmission
coefficient for the spin-flip case is always smaller than that for the
no spin-flip case for all non-zero values of coupling strength $G$. In
other words the reduction of amplitude of AB oscillations is stronger
for the spin-flip scattering case. We have verified the above
observation for other parameters in the problem.  Thus the presence of
spin-flipper reduces the AB-oscillations. This substantiates our claim 
of decoherence due to entanglement.

\begin{figure}
\protect\centerline{\epsfxsize=2.5in \epsfbox{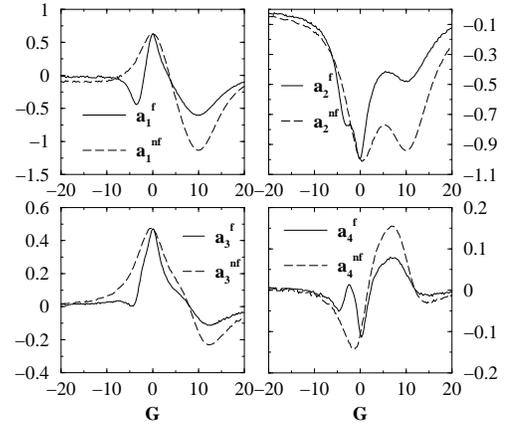}}
\caption{ Variation of $n^{th}$ harmonic component $a_n$ of the total
transmission coefficient with the interaction strength $G$ at $kl=1.0$. 
Dashed lines are for the no-flip case and solid lines are for the
flip case.}
\label{fourier}
\end{figure}

At this point we are inclined to think that the harmonic components of
the total transmission $T(\eta)$ in $\eta=\phi/\phi_0$ might be able
to shed more light on the issue. So, with the hope of extracting some
systematics we plot the $n^{th}$ harmonic component $a_n=\int_0^{2\pi}
T(\eta) \cos(n \eta) d \eta$ for $n=1,2,3...$ as a function of strength
$G$ for the spin-flip scattering as well as no spin-flip scattering
cases. The plots are shown in Fig. \ref{fourier} for first four
harmonic components.  As can be seen the harmonic components do not
show any systematics in the sense that the higher harmonic components
can dominate over the lower harmonic components at certain values of
strength $G$ for spin-flip scattering as well as no spin-flip
scattering cases. {\em Also, the} $n^{th}$ {\em harmonic component for
  spin-flip scattering could dominate over that of the no spin-flip
  scattering component}.  These features of the harmonic components
are manifestations of the multiple scattering nature of the transport
in such ballistic systems as against the observation of domination of
lowest harmonic component ($n=1$) in the case of transport in the
presence of evanescent modes \cite{ev_modes}. Guided by the naive
intuition mentioned earlier we would have expected the lowest harmonic
to dominate. This reiterates the important role played by the
reflection at the impurity site. We would like to emphasize that
irrespective of the behavior of the harmonic components (say for a
particular case $n^{th}$ harmonic component in the spin-flip case is
dominant over the same $n^{th}$ harmonic component for no-spin-flip
case) the AB-oscillations of the total transmission are always
suppressed in the spin-flip case.

\begin{figure}
\protect\centerline{\epsfxsize=2.5in \epsfbox{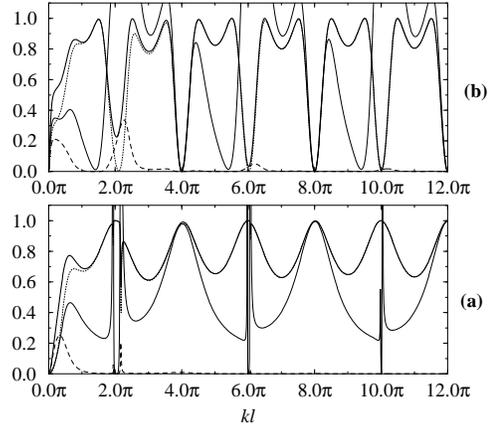}}
\caption{ The transmission spectrum of $T$(thick line), $T_u$(dotted line)
          and $T_d$(dashed line) for
          (a) $\eta=0.0$ $G=5.0$ and (b) $\eta=1.3$,$G=5.0$. Thin
          line shows the plot of the unnormalized electron probability
          $|\psi_3(l_3)|^2$ at the impurity site $x=l3$.}
\label{kspec}
\end{figure}

In order to make sure that nothing unusual happens at other energies
we study the $T$, $T_u$ and $T_d$ as functions of $kl$ for the case of
spin-flip scattering.  Figure \ref{kspec} reveals an interesting fact,
namely at $kl=2\pi+4n\pi,~n=0,1,2...$ the $T_d$ component vanishes
independent of the value of interaction strength $G$. In the $\eta\neq
0$ case this happens at $kl=4n\pi,~n=1,2,...$. At these values of the
incident wave-vectors the electron wave function at the impurity site
happens to be zero. As a result the electron does not interact with
the impurity at all and consequently there is no spin-flip scattering
at these energies.  However, these k-points are to be distinguished
from those at which although $T_d$ is zero but in addition $T_u=1$,
because at these resonant energies the restriction $T+R=1$ forces
$T_d,R_u$ and $R_d$ to be zero. 

In our separate study \cite{colin} we have analysed the same model in
regard to current magnification effect \cite{cme1,cme2,moskalets,choi}
which is also a purely quantum phenomenon. Against our naive intuition
we find that in some parameter regime spin-flip scattering (or
entanglement) enhances the magnitude of the current magnification as
opposed to the suppression of the AB-oscillations. According to the
general notion of dephasing or decoherence one expects all typical
features of quantum mechanical probability effects to be suppressed.
In full generality dephasing can be defined as the phenomenon by which
quantum mechanical systems behave as though they are described by
classical probability theory. Only the presence of inelastic
scattering or coupling the system to infinite environmental degrees of
freedom~(bath), leading to irreversible loss of phase memory, can
dephase both AB-oscillations and reduce current magnification effect
simultaneously. We believe that the suppression of some quantum
features and non-suppression other quantum effects is a characteristic
of entanglement and the absence of inelastic scattering. We expect the
same in other models based only on the notion of entanglement.
Moreover, let us emphasize that in our model the environment consists
of a single atom only.

In conclusion, we have studied in detail the nature of reduction of
AB-oscillations in mesoscopic ring in the presence of a spin-flipper
in one of its arms.  The presence of magnetic impurity makes the
polarized transmission coefficient asymmetric in flux reversal whereas
the total transmission coefficient is symmetric in line with the
theoretical expectations. We have also pointed out the limitations of
this entanglement based model in describing the phenomenon of dephasing
in quantum systems. Further case of spin-flipper with higher number of
internal states and spin-flippers in both arms of the ring are under
investigation.

\acknowledgments
One of us (DS) would like to thank Professor S. N. Behera for extending
hospitality at the Institute of Physics, Bhubaneswar.

%\begin{figure}[t]
%\protect\centerline{\epsfxsize=3in \epsfbox{amp1.eps}}
%\caption{Variation of Amplitude of AB oscillations with increasing strength
%$G$ of spin-flipper for the symmetrically placed flipper. $l_3/l=l_4/l=0.25$
%and $kl=1.0$.}
%\label{symm-imp}
%\end{figure}

\end{multicols}
\end{document}